\documentclass[aps,preprint,showpacs]{revtex4}
\usepackage[dvips]{graphicx}
\begin{document}
\title{Sine-Gordon model coupled with a free scalar field emergent
in the low-energy  phase dynamics of a mixture of pseudospin-$\frac{1}{2}$ Bose gases with
interspecies spin exchange}
\author{Li Ge}
\affiliation{Department of Physics and State Key Laboratory of
Surface Physics, Fudan University, Shanghai, 200433, China}
\author{Yu Shi}
\email{yushi@fudan.edu.cn}
\affiliation{Department of Physics and
State Key Laboratory of Surface Physics, Fudan University, Shanghai,
200433, China}

\begin{abstract}

Using the approach of low-energy effective field theory, the phase diagram is studied for a mixture of two species of pseudospin-$\frac{1}{2}$ Bose atoms with interspecies
spin-exchange.  There are four mean-field regimes on the parameter plane of $g_e$ and $g_z$, where $g_e$ is the interspecies spin-exchange interaction strength, while $g_z$ is the difference between the interaction strength of interspecies scattering without spin-exchange of equal spins and that of unequal spins. Two regimes, with $|g_z| > |g_e|$, correspond to  ground states with the total spins of the two species parallel or antiparallel along $z$ direction, and the low energy excitations are equivalent to those of two-component spinless Bosons.  The other two regimes, with  $|g_e| > |g_z|$, correspond to  ground states with the total spins of the two species parallel or antiparallel on  $xy$ plane, and the low energy excitations are described by a sine-Gordon model coupled with a free scalar field, where the effective fields are combinations of the phases of the original four Boson fields.  In (1+1)-dimension,  they are described by Kosterlitz-Thouless renormalization group (RG) equations, and there are three sectors in the phase plane of a scaling dimension and a dimensionless parameter proportional to the strength of the cosine interaction, both depending on the densities. The gaps of these elementary excitations are  experimental probes of the underlying many-body ground states.

\end{abstract}

\pacs{03.75.Mn, 11.10.Hi, 11.10.Kk}

\maketitle

\section{Introduction}

Sine-Gordon model is   important  in field theory and
statistical mechanics, and its renormalization group equations
define the Kosterlitz-Thouless universality class for a class of
(1+1)-dimensional quantum systems and two-dimensional classical
systems~\cite{gogolin,giamarchi}.  In this paper, we show that a sine-Gordon model coupled with a free scalar field emerges in the phase dynamics of a  mixture of two different species of pseudospin-$\frac{1}{2}$ Bose gases with interspecies spin-exchange interaction. Interestingly, here the scalar field described by the sine-Gordon model and the free scalar field are two different combinations of the phases of the four bosonic fields in this system.

A mixture of two different species of pseudospin-$\frac{1}{2}$ Bose gases with interspecies spin-exchange interaction exhibits novel features beyond a single species of spinor Bose gas as well as a mixture of two species without interspecies spin exchange~\cite{shi0,shi1,shi2,shi3,wang,wu,wu2,ge}.   In this model, each atom has an internal
degree of freedom represented as a pseudospin with
basis states $|\uparrow\rangle$ and $|\downarrow\rangle$, while there are two species of atoms, with the atom number of each species conserved. It can be described by the following Hamiltonian density,
\begin{equation}
\begin{array}{rl}
\displaystyle
\mathcal{H}=&
\displaystyle  \sum_{\alpha\sigma} \psi_{\alpha\sigma}^\dagger (-\frac{1}{2m_\alpha}\nabla^2 +   V) \psi_{\alpha\sigma} + \frac{1}{2}\sum_{\alpha\sigma\sigma^{'}}{
g^{(\alpha\alpha)}_{\sigma\sigma^{'}}
|\psi_{\alpha\sigma}|^{2}|\psi_{\alpha\sigma^{'}}|^{2}} \\
\displaystyle
&
\displaystyle +\sum_{\sigma\sigma^{'}}{g^{(ab)}_{\sigma\sigma^{'}}|\psi_{a\sigma}|^{2}
|\psi_{b\sigma^{'}}|^{2}}+
g_{e}(\psi_{a\uparrow}^{\dagger}
\psi_{a\downarrow}\psi_{b\downarrow}^{\dagger}\psi_{b\uparrow}+
\psi_{a\downarrow}^{\dagger}\psi_{a\uparrow}
\psi_{b\uparrow}^{\dagger}\psi_{b\downarrow}),
\end{array}
\end{equation}
where $\alpha=a, b$ represents the two species and $\sigma=\uparrow, \downarrow$.
$V=V(x)$ is the external potential, $g^{(\alpha\alpha)}_{\sigma\sigma^{'}}$, $g^{(ab)}_{\sigma\sigma^{'}}$ and $g_{e}$ are the interaction strengths for intraspecies scattering, interspecies scattering without spin exchange, and interspecies spin-exchange scattering respectively,  proportional to the corresponding scattering lengths.  For pseudospin-$\frac{1}{2}$ atoms, intraspecies scattering strengths with and without spin-exchange are the same~\cite{leggett}.  For simplicity, we assume
$g^{(\alpha\alpha)}_{\sigma{\sigma}'}=g_{\alpha}$ for any $\sigma$ and $\sigma'$, $g^{(ab)}_{\uparrow\uparrow}=g^{(ab)}_{\downarrow\downarrow}=g_s$ and  $g^{(ab)}_{\uparrow\downarrow}=g^{(ab)}_{\downarrow\uparrow}=g_d$. We define ${\cal S}_{\alpha i}(x)= \Psi^{\dagger}_{\alpha}{s}^{i}\Psi_{\alpha}$,  where   $\Psi_{\alpha}(x) \equiv (\psi_{\uparrow}(x),\psi_{\downarrow}(x))^T$, $s^{i}=\tau_i/2$,  $\tau_i$ being the Pauli matrix,  $(i=x,y,z)$. Then   ${\cal H}$  can be rewritten as
\begin{equation}
\begin{array}{rl}
\displaystyle
{\cal H} =  &
\displaystyle \sum_{\alpha} \Psi_{\alpha}^\dagger (-\frac{1}{2m_\alpha}\nabla^2 +   V ) \Psi_{\alpha} +
\frac{g_a}{2}|\Psi_a|^4+\frac{g_b}{2}|\Psi_b|^4
+\frac{g_{ab}}{2}|\Psi_a|^2|\Psi_b|^2
\\
\displaystyle &
\displaystyle
+2g_{e}({\cal S}_{ax}{\cal S}_{bx}+ {\cal S}_{ay}{\cal S}_{by})+2g_z {\cal S}_{az}{\cal S}_{bz}, \label{hamil}
\end{array}
\end{equation}
where $g_{ab} \equiv g_s+g_d$, $g_z \equiv g_s-g_d$.
It can be seen that $g_a$, $g_b$, and $g_{ab}$ characterize the usual
density-density interactions, while $g_z$ and $g_e$ characterize the  spin coupling between the two species.

We make the presumption that  $g_a>0$, $g_b>0$ and $4g_ag_b>g_{ab}^2$, which is needed for
the stability of the system and can be naturally
satisfied in reality~\cite{ge}.  We study the phase diagram in the space of the parameters  $g_e$
and $g_z$, by using the  approach of low energy
effective field theory. The regime of $g_e>g_z>0$ has been discussed
previously for higher dimensions, i.e. when the phase fluctuation is suppressed such that the cosine of a phase variable can be approximated up to second order~\cite{ge}.  In this paper, we first extend the discussion to
other parameter regimes. In the regime of  $g_z>|g_e|$, the ground state is with the total spins of the two species antiparallel  along $z$ direction. In the regime of $g_z < -|g_e|$, the ground  state is with the total spins of the two species parallel along $z$ direction.  In the regime of   $g_e>|g_z|$, the ground   state is with the total spins of the two species antiparallel on  $xy$ direction. In the regime of    $g_e < -|g_z|$, the ground  state is with the total spins of the two species  parallel on $xy$ direction. Then we focus  on the case of  $|g_e|>|g_z|$ in
(1+1)-dimension. Without approximating the cosine interaction term, the low energy excitations can be described by a sine-Gordon model coupled with a
free scalar field, both fields being combinations of the phases of the
original four boson field. It turns out that for given $g_e$ and $g_z$ with  $|g_e| > |g_z|$,  there are three
phases according to a scaling dimension and a dimensionless parameter proportional to $|g_e|$. Both these two parameters depend on the densities of the two species.

\section{Phase diagram on $g_e-g_z$ parameter plane}

There is a symmetry between parameter points $(g_e, g_z)$ and $(-g_e, g_z)$. Consider the transformation $\psi'_{a\uparrow} \equiv -\psi_{a\uparrow}$,
$\psi'_{a\downarrow} \equiv \psi_{a\downarrow}$, $\psi'_{b\uparrow}
\equiv \psi_{b\uparrow}$, $\psi'_{b\downarrow} \equiv
\psi_{b\downarrow}$. The Hamiltonian density in terms
of the primmed operators in the parameter point  $(g_e, g_z)$  has the same form as the Hamiltonian density in terms
of the unprimed ones in the
parameter point $(-g_e, g_z)$.

Now consider the case of $g_z>|g_e|$. Then from the Hamiltonian density (\ref{hamil}), it is
easy to see that in the ground state,  $\mathbf{{\cal S}}_a$ and
$\mathbf{{\cal S}}_b$ must align oppositely in the $z$ direction. We
choose the mean field values in the  ground state to be with
$\psi_{a\uparrow}^0=\sqrt{n_a}$, $\psi_{a\downarrow}^0=0$,
$\psi_{b\uparrow}^0=0$, $\psi_{b\downarrow}^0=\sqrt{n_b}$ so that
$\mathbf{\cal S}_a=\frac{n_a}{2} \hat{z}$ and  $\mathbf{{\cal S}}_b=-\frac{n_b}{2} \hat{z}$, where $\hat{z}$ is the unit vector in the $z$ direction, $n_\alpha$ is the total density of species $\alpha$, ($\alpha =a, b$). The low energy dynamics is dominated by the fluctuations of $\psi_{a\uparrow}$ and $\psi_{b\downarrow}$,  as the fluctuations of  $\psi_{a\downarrow}$ and $\psi_{b\uparrow}$, whose mean field values are zero, must be of the amplitudes rather than the phases, and thus increase the energy.

Similarly, in the case of $g_z <-|g_e|$, the ground state is that with   $\mathbf{{\cal S}}_a$ and
$\mathbf{{\cal S}}_b$ parallel in the $z$ direction. We can
choose the mean field values in the  ground state to be with
$\psi_{a\uparrow}^0=\sqrt{n_a}$, $\psi_{a\downarrow}^0=0$,
$\psi_{b\uparrow}^0=\sqrt{n_b}$, $\psi_{b\downarrow}^0=0$,  so that
$\mathbf{\cal S}_a=\frac{n_a}{2} \hat{z}$ and  $\mathbf{{\cal
    S}}_b=\frac{n_b}{2} \hat{z}$. The low energy dynamics is
  dominated by the fluctuations of $\psi_{a\uparrow}$ and
  $\psi_{b\uparrow}$,  as the fluctuations of  $\psi_{a\downarrow}$
  and $\psi_{b\downarrow}$, whose mean field values are zero, must be
  of the amplitudes and thus increase the energy.

In these two cases, which can be represented in a unified form as $|g_z| > |g_e|$,  the system behaves like a mixture of two species of spinless Boson gases,  with the effective Hamiltonian   density
\begin{equation}
{\cal H} =
\displaystyle \sum_{\alpha} \Psi_{\alpha}^\dagger (-\frac{1}{2m_\alpha}\nabla^2 +   V ) \Psi_{\alpha} +
\frac{g_a}{2}|\Psi_a|^4+\frac{g_b}{2}|\Psi_b|^4
+\frac{g_{ab}-|g_z|}{2}|\Psi_a|^2|\Psi_b|^2, \label{h01}
\end{equation}
whose  excitation spectra are \cite{pethick}
\begin{equation}
\omega^2=\frac{1}{2}(\varepsilon_a^2+\varepsilon_b^2)\pm
\frac{1}{2}\sqrt{(\varepsilon_a^2-\varepsilon_b^2)^2+4E_aE_bn_an_b(g_{ab}-|g_z|)^2}, \label{spec01}
\end{equation}
where we have introduced
\begin{equation}
\varepsilon_\alpha^2 \equiv  E_\alpha (2g_\alpha n_\alpha + E_\alpha),
\end{equation}
with $\alpha=a, b$, and $E_\alpha=\frac{k^2}{2m_\alpha}$ being the
free particle energy of species $\alpha$. All spectra are gapless, as in the usual case of phonon-like Goldstone modes. That is, $\omega \rightarrow 0$ when $k \rightarrow 0$.

In the case of $g_e > |g_z|$, which has been discussed previously~\cite{ge},  the ground state is that with   $\mathbf{{\cal S}}_a$ and
$\mathbf{{\cal S}}_b$ antiparallel on  the $xy$ plane.  One can choose
the ground state to be with  $\psi_{a\uparrow}^0=\psi_{a\downarrow}^0=\sqrt{n_a/2}$,
$\psi_{b\uparrow}^0=-\psi_{b\downarrow}^0=\sqrt{n_b/2}$.

Similarly, in the case of  $g_e < -|g_z|$, the ground state is that with   $\mathbf{{\cal S}}_a$ and
$\mathbf{{\cal S}}_b$ parallel on  the $xy$ plane.  One can choose
the ground state to be with  $\psi_{a\uparrow}^0=\psi_{a\downarrow}^0=\sqrt{n_a/2}$,
$\psi_{b\uparrow}^0= \psi_{b\downarrow}^0=\sqrt{n_b/2}$.

In the latter two cases, which can be represented in a unified form as $|g_e| > |g_z|$,  the effective Lagrangian describing the phase fluctuations is, with $|g_e|$ replacing $g_e$ in the result  for  $g_e > g_z > 0 $ \cite{ge}, that is,

  \begin{equation}
  \mathcal{L}_{eff}=\frac{1}{2} (\partial_t\Gamma^T) {A}^{-1}(\partial_t\Gamma)
  -\frac{1}{2}(\nabla{\Gamma}^T) M^{-1}(\nabla{\Gamma})
  +\frac{|g_e|}{2}n_a n_b \cos{(2\gamma_4)} \label{effect}
  \end{equation}
  where
   \begin{equation}
   \Gamma=\left(
           \begin{array}{c}
             \gamma_1 \\
             \gamma_2\\
             \gamma_3 \\
             \gamma_4 \\
           \end{array}
         \right)=\left(
           \begin{array}{cccc}
             \frac{1}{\sqrt{2}} & \frac{1}{\sqrt{2}} & 0 & 0 \\
             0 & 0 & \frac{1}{\sqrt{2}} & \frac{1}{\sqrt{2}} \\
             \frac{1}{2} & -\frac{1}{2} & \frac{1}{2} & -\frac{1}{2} \\
             \frac{1}{2} & -\frac{1}{2} & -\frac{1}{2} & \frac{1}{2} \\
           \end{array}
         \right)\left(
                   \begin{array}{c}
                     \Phi_{a\uparrow} \\
                     \Phi_{a\downarrow} \\
                     \Phi_{b\uparrow} \\
                     \Phi_{b\downarrow} \\
                   \end{array}
                 \right),
  \end{equation}
 with $\Phi_{\alpha \sigma}$ being the phase of $\psi_{\alpha \sigma}$,
 \begin{equation}
 A \equiv \left(
                                        \begin{array}{cccc}
                                          2g_a & g_{ab}-|g_e| & 0 & 0 \\
                                          g_{ab}-|g_e| & 2g_b & 0 & 0 \\
                                          0 & 0 & |g_e|\eta_{+}+g_z & |g_e|\eta_{-} \\
                                          0 & 0 & |g_e|\eta_{-} & |g_e|\eta_{+}-g_z \\
                                        \end{array}
                                      \right),
  \end{equation}
  with  $$\eta_{\pm } \equiv \frac{1}{2}(\frac{n_b}{n_a} \pm \frac{n_a}{n_b}),$$
  \begin{equation}
 {M}^{-1} \equiv \frac{1}{2} \left(
               \begin{array}{cccc}
                 \frac{n_a}{m_a} & 0 & 0 & 0 \\
                 0 & \frac{n_b}{m_b} & 0 & 0 \\
                 0 & 0 & \xi_+ & \xi_- \\
                 0 & 0 & \xi_- & \xi_+ \\
               \end{array}
             \right),
  \end{equation}
 with $$\xi_{\pm } \equiv \frac{1}{2}(\frac{n_a}{m_a}\pm \frac{n_b}{m_b}).$$

In (3+1)-dimension or (2+1)-dimension,  the fluctuation of $\gamma_4$ is largely suppressed and we can make the approximation that  $\cos(2\gamma_4) \approx 1-2\gamma_4^2$, subsequently, the four excitation spectra can be obtained as \cite{ge},
\begin{equation}
\omega^2_{I,II}=\frac{k^2}{2}\bigg[\frac{g_an_a}{m_a}+\frac{g_bn_b}{m_b}
\mp \sqrt{(\frac{g_an_a}{m_a}-\frac{g_bn_b}{m_b})^2+
\frac{(g_{ab}-|g_e|)^2n_an_b}{m_am_b}}\bigg],  \label{excitation1}
\end{equation}
\begin{equation}
\omega^2_{III,IV}= \frac{1}{2}\bigg[ Bk^2
+\Delta^2
\mp \sqrt{Ck^4+Dk^2+\Delta^4 }\bigg],  \label{excitation2}
\end{equation}
where
$\Delta^2 =  |g_e^2 (\frac{n_b}{n_a}+\frac{n_a}{n_b})-2|g_e|g_z|n_an_b,$
$B\equiv \frac{|g_e|}{2}(\frac{n_b}{m_a}+\frac{n_a}{m_b})$,
$C\equiv \frac{g_e^2}{4}(\frac{n_b}{m_a}-\frac{n_a}{m_b})^2+g_z^2\frac{n_an_b}{m_am_b}$,
$D\equiv |g_e|n_an_b[g_e^2 (\frac{n_b}{m_a}-\frac{n_a}{m_b})(\frac{n_b}
{n_a}-\frac{n_a}{n_b})-2|g_e|g_z(\frac{n_b}{m_a}+\frac{n_a}{m_b})+2g_z^2(\frac{n_a}{m_a}+\frac{n_b}{m_b})]$.
Under the conditions $g_a>0$, $4g_ag_b>g_{ab}^2$ and $|g_e|>|g_z|$,  all these excitations have real energies for any $k$, guaranteeing  the stability of the ground state.

It can be seen that $\omega_{IV}$ has a gap $\Delta$ while the other three excitations are gapless. That is, as $k \rightarrow 0$, $\omega_{I,II,III} \rightarrow 0$, but  $\omega_{IV} \rightarrow \Delta$.

\begin{figure}
\begin{center}
\includegraphics{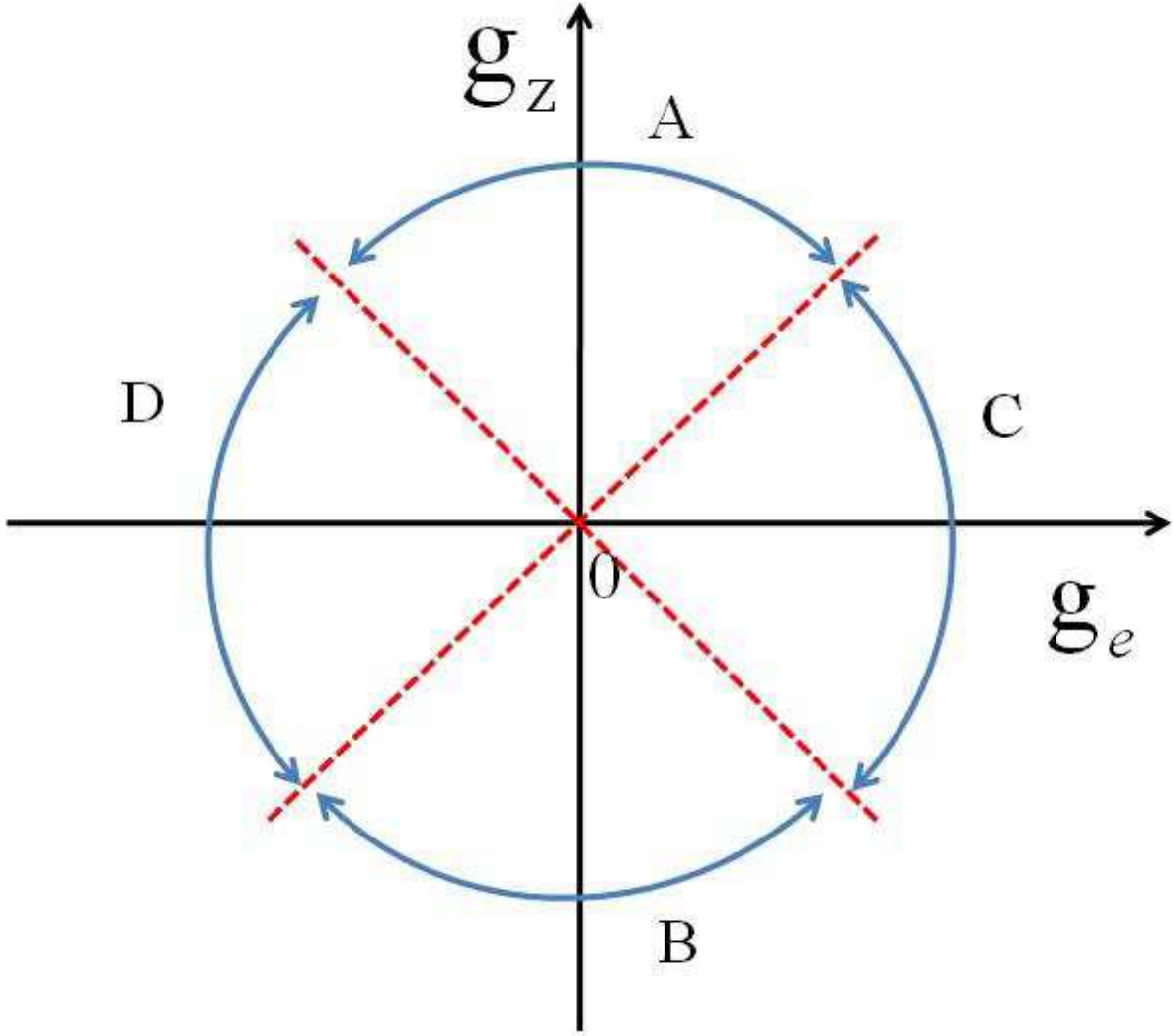}
\caption{ \label{pd} Mean-field phase diagram on $g_e-g_z$ phase plane in (3+1)-dimension. In regime A,   $g_z>|g_e|$, the ground state is $\psi_{a\uparrow}^0=\sqrt{n_a}$, $\psi_{a\downarrow}^0=0$, $\psi_{b\uparrow}^0=0$, $\psi_{b\downarrow}^0=\sqrt{n_b}$.  In regime B, $g_z < -|g_e|$,  the ground state is $\psi_{a\uparrow}^0=\sqrt{n_a}$, $\psi_{a\downarrow}^0=0$, $\psi_{b\uparrow}^0=\sqrt{n_b}$, $\psi_{b\downarrow}^0=0$.   In regime C,  $g_e > |g_z|$, the ground state is $\psi_{a\uparrow}^0=\psi_{a\downarrow}^0=\sqrt{n_a/2}$,  $\psi_{b\uparrow}^0=-\psi_{b\downarrow}^0=\sqrt{n_b/2}$.  In regime D, $g_e<-|g_z|$, the ground state is    $\psi_{a\uparrow}^0=\psi_{a\downarrow}^0=\sqrt{n_a/2}$,  $\psi_{b\uparrow}^0=\psi_{b\downarrow}^0=\sqrt{n_b/2}$.  \label{f1}  }
\end{center}
\end{figure}

Therefore in (3+1)-dimension,   we obtain the mean-field phase diagram as shown in Fig.~\ref{pd}.

\section{Renormalization group analysis in the case of $|g_e| > |g_z|$ in (1+1)-dimension }

Reconsider the case of  $|g_e| > |g_z|$. In (1+1)-dimension, the fluctuation is important and we must take into account the whole effect of $\cos(2\gamma_4)$ term, which is the only interaction term in  ${\cal L}_{eff}$, where  $\gamma_1$ and $\gamma_2$ are both free fields and not coupled to $\gamma_4$.  Hence we can focus on the sine-Gordon field $\gamma_4$ coupled to a free scalar field $\gamma_3$,
\begin{eqnarray}
\mathcal{L}=\frac{1}{2(g_e^2-g_z^2)} \left(
                         \begin{array}{cc}
                           \partial_t \gamma_3 & \partial_t \gamma_4 \\
                         \end{array}
                       \right) \left(
                                 \begin{array}{cc}
                                   |g_e|\eta_{+}-g_z & -|g_e|\eta_{-} \\
                                   -|g_e|\eta_{-} & |g_e|\eta_{+}+g_z \\
                                 \end{array}
                               \right)
                               \left(
                                             \begin{array}{c}
                                               \partial_t \gamma_3 \\
                                               \partial_t \gamma_4 \\
                                             \end{array}
                                           \right)  \nonumber \\
                                          -\frac{1}{4} \left(
                         \begin{array}{cc}
                           \partial_x \gamma_3 & \partial_x \gamma_4 \\
                         \end{array}
                       \right) \left(
                                 \begin{array}{cc}
                                   \xi_+ & \xi_- \\
                                   \xi_- & \xi_+ \\
                                 \end{array}
                               \right) \left(
                                             \begin{array}{c}
                                               \partial_x \gamma_3 \\
                                               \partial_x \gamma_4 \\
                                             \end{array}
                       \right)+\frac{|g_e|}{2}n_a n_b \cos{(2\gamma_4)}. \label{reduce}
\end{eqnarray}

We now    make a  renormalization group (RG)  analysis of the cosine interaction term due to spin exchange in (1+1)-dimension, by following the approach in \cite{gogolin}. It turns out that it still belongs to the Kosterlitz-Thouless universality class. But the novelty is that the scalar fields  are now combinations of the phases of the original bosonic fields.

Let us define  $x^0 \equiv vt$, with $v \equiv \sqrt{\frac{(g_e^2-g_z^2)\xi_+}{2 (|g_e|\eta_{+}+g_z)}}$, and  two dimensionless field variables corresponding to $\gamma_3$ and $\gamma_4$
$$\varphi \equiv [\frac{\xi_+(|g_e|\eta_{+}+g_z)}{2(g_e^2-g_z^2)}]^{\frac{1}{4}}\gamma_3,$$
$$\chi \equiv[\frac{\xi_+(|g_e|\eta_{+}+g_z)}{2(g_e^2-g_z^2)}]^{\frac{1}{4}}\gamma_4.$$
Then the action can be written as
\begin{equation}
S=S_0+S_I,
\end{equation}
where $S_0 \equiv \int \mathcal{L}_0 d^2x$,    $S_I \equiv \int \mathcal{L}_I d^2x$, with
\begin{equation}
\mathcal{L}_0=\frac{1}{2} \left(
                         \begin{array}{cc}
                           \partial_0 \varphi & \partial_0 \chi \\
                         \end{array}
                       \right) \left(
                                 \begin{array}{cc}
                                   f_1 & f_2 \\
                                   f_2 & 1 \\
                                 \end{array}
                               \right)
                               \left(
                                             \begin{array}{c}
                                               \partial_0 \varphi \\
                                               \partial_0 \chi \\
                                             \end{array}
                                           \right)
                                          +\frac{1}{2} \left(
                         \begin{array}{cc}
                           \partial_1 \varphi & \partial_1 \chi \\
                         \end{array}
                       \right) \left(
                                 \begin{array}{cc}
                                   1 & p \\
                                   p & 1 \\
                                 \end{array}
                               \right) \left(
                                             \begin{array}{c}
                                               \partial_1 \varphi \\
                                               \partial_1 \chi \\
                                             \end{array}
                       \right),  \label{couple}
\end{equation}
\begin{equation}
\mathcal{L}_I=\frac{\lambda}{a^2}\cos{(\beta\chi)}, \label{li}
\end{equation}
where $$f_1 \equiv \frac{|g_e|\eta_{+}-g_z}{|g_e|\eta_{+}+g_z},$$ $$f_2 \equiv -\frac{|g_e|\eta_{-}}{|g_e|\eta_{+}+g_z},$$
$$p \equiv  \frac{\xi_-}{\xi_+},$$  $$\beta \equiv 2 [\frac{\xi_+(|g_e|\eta_{+}+g_z)}{2(g_e^2-g_z^2)}]^{-\frac{1}{4}},$$
$$\lambda \equiv  \frac{|g_e|n_an_b a^2}{2v},$$ $a$ is the short range cut-off, which is the coherence or healing length, and can be estimated to be $\hbar[m_b(g_a n_a^2 + g_b n_b^2 + (g_s+g_d-g_e)n_a n_b)]^{-1/2}$, assuming $m_a \geq m_b$.  $\lambda$ is  dimensionless.
From $\mathcal{L}_0$  the free propagator of $\chi$ is obtained as
\begin{equation}
\mathcal{G}_\chi(\mathbf{k})=\frac{f_1k_0^2+k_1^2}{k^2(f_1k_0^2+k_1^2)-(f_2k_0^2+pk_1^2)^2}=\frac{1}{k^2} \frac{f_1\cos^2\theta+\sin^2\theta}{f_1\cos^2\theta+\sin^2\theta-
(f_2\cos^2\theta+p\sin^2\theta)^2},
\end{equation}
where $k^2=k_0^2+k_1^2$, $\tan \theta=\frac{k_1}{k_0}$.

Since the interaction term does not involve $\varphi$ field, we only need  to split $\chi$ into the fast   and slow components,
\begin{equation}
\chi_\Lambda(x)={\chi_{\Lambda'}(x)+h(x)},
\end{equation}
where
\begin{equation}
\chi_{\Lambda'}(x) \equiv
\sum_{k<\Lambda'}e^{ikx}\chi_k,
\end{equation}
\begin{equation}
h(x) \equiv \sum_{\Lambda'<k<\Lambda}e^{ikx}\chi_k,
\end{equation}
 $\Lambda=\frac{1}{a}$ is the momentum cut-off,
 $\Lambda'=\Lambda-d\Lambda$. The partition function can be written as~\cite{gogolin}
\begin{eqnarray}
Z_\Lambda&=&\int \mathcal{D}\varphi \mathcal{D}\chi_{\Lambda'}\mathcal{D}h e^{-S_0[\varphi, \chi_{\Lambda'}]-S_0[h]-S_I[\chi_{\Lambda'}+h]} \\
&=&Z_h\int \mathcal{D}\varphi \mathcal{D}\chi_{\Lambda'} e^{-S_0[\varphi, \chi_{\Lambda'}]}\langle e^{-S_I[\chi_{\Lambda'}+h]}\rangle_h,
\end{eqnarray}
where $Z_h=\int \mathcal{D}h e^{-S_0[h]}$, $\langle ... \rangle_h$ means taking average over the fast components of $\chi$.

The effective action is thus
\begin{eqnarray}
S[\varphi, \chi_{\Lambda'}]&=&S_0[\varphi, \chi_{\Lambda'}]-\ln \langle e^{-S_I[\chi_{\Lambda'}+h]}\rangle_h \nonumber \\
& \approx & S_0[\varphi, \chi_{\Lambda'}]+ \langle S_I[\chi_{\Lambda'}+h]\rangle_h-\frac{1}{2}\Big (\langle S^2_I[\chi_{\Lambda'}+h]\rangle_h-\langle S_I[\chi_{\Lambda'}+h]\rangle^2_h \Big),
\end{eqnarray}
which  allows us to calculate the RG flows of $\lambda$ and $\beta$. We have
\begin{equation}
\langle h(x)h(0) \rangle_h=\int_{\Lambda'<k<\Lambda}\frac{d^2k}{(2\pi)^2} \mathcal{G}_\chi(\mathbf{k}) e^{i\mathbf{k}\cdot \mathbf{x}}
=\frac{\kappa(\Lambda r)}{2\pi}dl,
\end{equation}
where $$dl\equiv\frac{d\Lambda}{\Lambda},$$ $r \equiv |\mathbf{x}|$,
\begin{equation}
\kappa(\Lambda r) \equiv \int \frac{d\theta}{2\pi}\frac{f_1\cos^2\theta+
\sin^2\theta}{f_1\cos^2\theta+\sin^2\theta-(f_2\cos^2\theta+p\sin^2\theta)^2}e^{i\Lambda r \cos\theta}, \label{correlator}
\end{equation}
which is a measure of correlation of fluctuations.

Then
\begin{equation}
\langle e^{i\beta h(x)}\rangle_h=e^{-\frac{1}{2}\beta^2\langle h^2(x) \rangle_h}=1-\frac{\beta^2\kappa(0)}{4\pi}dl\equiv 1-D dl.  \label{scaling}
\end{equation}
It is seen that the coupling between $\varphi$ and $\chi$ modifies the scaling dimension of $e^{i\beta\chi}$ from $D_0=\frac{\beta^2}{4\pi}$~\cite{gogolin} to
\begin{equation}
D=\frac{\beta^2\kappa(0)}{4\pi}, \label{d}
\end{equation}
which means the interaction term $\cos(\beta\chi)$ is less relevant than that in the pure SG model, since $\kappa(0)>1$.

Now we can obtain  the renormalized action. Following ~\cite{gogolin}, we have
\begin{equation}
\langle S_I[\chi_{\Lambda'}+h]\rangle_h=\lambda (1- D dl)\int \frac{d^2x}{a^2}\cos(\beta\chi_{\Lambda'}), \label{first}
\end{equation}
\begin{equation}
\langle S^2_I\rangle_h-\langle S_I\rangle^2_h=-\alpha\lambda^2D^2_0dl\int d^2x (\nabla\chi_{\Lambda'})^2,  \label{second}
\end{equation}
where 
\begin{equation} 
\alpha=\int^\infty_0 \frac{dz}{2\pi}z^3\kappa(z).
\end{equation}

Therefore, by rescaling $\Lambda'\rightarrow\Lambda$, the purely $\chi$-dependent  part of the action is renormalized to
\begin{equation}
 S[\chi]=\frac{1}{2}(1+\alpha\lambda^2D^2_0dl) \int d^2x (\nabla\chi)^2+\lambda[1+(2-Ddl)]\int \frac{d^2x}{a^2}\cos(\beta\chi). \label{action}
\end{equation}
The overall factor in front of the Gaussian part of the action requires a renormalization of the field $\chi$, as well as the parameter $\beta$  such that  $\beta\chi$ is invariant. That is,
\begin{equation}
\begin{array}{rcl}
 \chi(x)& \rightarrow&  (1+\alpha\lambda^2D^2_0dl)^{1/2}\chi(x),    \\
 \beta & \rightarrow&  (1+\alpha\lambda^2D^2_0dl)^{-1/2}\beta,
 \end{array}
 \label{renormalize}
\end{equation}
which gives  RG flows of $\lambda$ and $\beta$.

There is another field $\varphi$ coupled to $\chi$. The coupling terms are those proportional to $f_2$ and $p$, respectively, in (\ref{couple}). The above renormalization of $\chi$ and $\beta$ lead to the renormaliztion of $f_2$ and $p$. As in the pure SG model, we introduce $$t \equiv D-2,$$ and assume $t$ and $\lambda$ are small, as the parameter point of $t=0$ and $\lambda=0$ is a fixed point of the RG flows. Then  to  the order of $\lambda^2$, the RG equations read
\begin{equation}
\begin{array}{rl}
& \frac{dt}{dl}=-\frac{8\alpha\lambda^2}{\kappa^2(0)},  \\
& \frac{d\lambda}{dl}=-\lambda t, \\
& \frac{df_2}{dl}=-\frac{2\alpha\lambda^2}{\kappa^2(0)}f_2, \\
& \frac{dp}{dl}=-\frac{2\alpha\lambda^2}{\kappa^2(0)}p.
\end{array}
  \label{RG}
\end{equation}

\section{Phase diagram on the $t-y$ parameter plane and the mass gap in (1+1)-dimension}

\begin{figure}
\begin{center}
\includegraphics[0pt,0pt][152pt,108pt]{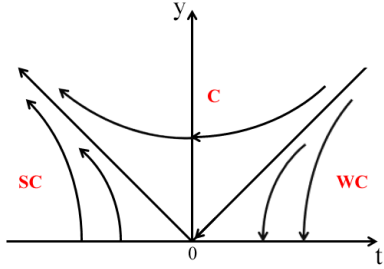}
\caption{Phase diagram of our emergent sine-Gordon model coupled
with a free field in (1+1)-dimension, with $y \equiv
\frac{\sqrt{8\alpha}}{\kappa(0)}\lambda$,  $t\equiv D-2$, $D$ is a
scaling dimension. There are two separatrices $t=\pm y$ that divide
the phase plane into three sectors: (1) $t\geq y$, the weak coupling
(WC) sector; (2) $ |t|<y$, the crossover (C) sector; (3) $t\leq -y$,
the strong coupling (SC) sector. The RG flows are similar to the pure
sine-Gordon model.   \label{fig2} }
\end{center}
\end{figure}

From the RG equations, one obtains \begin{equation}
d(t^2)-\frac{8\alpha}{\kappa^2(0)}d(\lambda^2)=0,
\end{equation}
which is similar to the equation in the pure SG model~\cite{gogolin}, except that $\alpha$ and $\kappa$ are not constant here. Moreover, $\frac{d\kappa(0)}{dl}$ and $\frac{d\alpha}{dl}$ are both proportional to $\lambda^2$. Hence to the order of $\lambda^3$, one can  replace $\frac{8\alpha}{\kappa^2(0)}d(\lambda^2)$ as $d(\frac{8\alpha}{\kappa^2(0)}\lambda^2)$. Thus we arrive at the following equation,
\begin{equation} \label{flow}
t^2- y^2  = \mu^2, \label{rge}
\end{equation}
where
\begin{equation}
y \equiv \frac{\sqrt{8\alpha}}{\kappa(0)}\lambda, \label{y}
\end{equation}
$\mu$ represents a constant. For given $g_e$ and $g_z$ with $|g_e| > |g_z|$,  Equation (\ref{rge})  determines the phase diagram  of the model on the plane $(t, y)$ in the regime where $t$ and $y$ are small, as schematically shown in Fig.~\ref{fig2}.

It can be seen that the   $(t, y)$  phase space is divided to three sectors, namely, weak coupling, strong coupling  and crossover sectors.
In the weak coupling sector, the effective theory scales to a Gaussian model, $y(l)\rightarrow 0$ as $l\rightarrow \infty$,  and the spectrum is massless, while in the crossover and strong coupling  sectors, the coupling constants flow away from the Gaussian fixed line and the spectrum has a mass gap.

Note that both $t$ and $y$ depend not only on the interspecies spin-exchange coupling, but also  on the densities. Consequently  the phase is dependent   not only on the interaction strengths,  but also on the densities of the two species, which can be easily adjusted in experiments. To illustrate this explicitly, let us set $n_a=n_b=n$ so that $f_1=\frac{|g_e|-g_z}{|g_e|+g_z}$, $f_2=0$, $p=\frac{m_b-m_a}{m_b+m_a}$ and  thus $\kappa$ and $\alpha$ as well,  are all independent of $n$,  while $\beta\propto n^{-\frac{1}{4}}$ and $\lambda\propto n^{\frac{3}{2}}$. Therefore $D \propto n^{-\frac{1}{2}}$ and $ y \propto n^{\frac{3}{2}}$.  Then according to Fig.~\ref{fig2}, for small enough $n$,   the system is in the weak coupling phase. For large enough $n$,   the system is in the strong coupling phase. Therefore,  following the change of density,  the system goes through phase transitions.

The mass gap in the crossover and strong coupling  sectors can be qualitatively obtained. In the strong coupling sector, $\mu$ is real, while in the crossover  sector, $\mu$ is purely imaginary. Following ~\cite{gogolin}, it can be found that  the mass gap in the strong coupling  sector is
\begin{equation}
 M=\left\{
 \begin{array}{ll}
 \Lambda(\frac{y_0}{t_0})^{1/t^0}, & -t_0\gg y_0, \\
 \Lambda \exp(-1/y_0), &  \mu\ll |y_0|,
 \end{array}
 \right.
 \end{equation}
where the subscript ``0'' means the bare values, that is, the values measured in experiments,
while in the crossover  sector, the mass gap is
\begin{equation}
M=\Lambda \exp(-\pi/2|\mu|),
\end{equation}
with $|\mu|\gg|t_0|$.

The scaling behavior of the mass gap  may be observed in experiments. Taking $M=\Lambda \exp(-1/y_0)$ as an example. If $n_a=n_b=n$,  we have $y_0=\frac{\sqrt{8\alpha}}{\kappa(0)}\lambda\sim n^{\frac{3}{2}}$, then the relation between $M$ and $n$ may be investigated.

The last two  of the RG equations (\ref{RG}) determine the RG flows of the couplings between the fields $\varphi$ and $\chi$.  Since $\alpha>0$, it is easy to find that $f_2=0$, $p=0$ is the only stable fixed point of the two equations, namely, whatever the initial values of $f_2$ and $p$ are, they inevitably flow to $0$. Moreover, the larger $\lambda$ is,  the more rapidly $f_2$ and $p$ flow to $0$. It is like that its strong self interaction ``traps'' the field $\chi$ and separate it from $\varphi$. If the bare  value of $\lambda$ were $0$, there would be no RG flows of $f_2$ and $p$.

Also note that if we diagonalize ${\cal L}_0$ in (\ref{couple}),  then ${\cal L}_I$ in (\ref{li}) becomes a cosine term of cosine of a linear combination of the two fields, of which the RG analysis is quite difficult. Hence we use the above approach instead.

The elementary excitations studied here  can be experimentally measured by using the Bragg spectroscopy.
The gap in a collective mode is a novel feature absent in the BEC  mixtures previously studied.   The two key parameters $g_e$ and  $g_z$  both originate from the interspecies spin-dependent scattering, thus they are roughly of the same order of magnitude.  We expect that  the excitation  gaps  can be detected in experiments and are indications of the underlying many-body ground states.

\section{Summary  \label{summary} }

We have developed a low energy effective theory for a mixture of two species of pseudospin-$\frac{1}{2}$ Bose gases and explore the phase transitions in the space of the parameters $g_e$ and $g_z$, where $g_e$ is the  interspecies spin-exchange interaction strength, while  $g_z$ is the difference between the strengths of equal-spin and unequal-spin interspecies interaction without spin exchange. The phase diagram on the plane of parameters $g_e$ and $g_z$ is shown in Fig~\ref{f1}. In the regime of $|g_z| > |g_e|$, the system is effectively described by a two component model, and the excitation spectra are gapless. In the regime of $|g_e| > |g_z|$, the system is described by a four effective fields, which are combinations of the phases of the four original boson fields.  There is a cosine interaction term of one of the effective field, which can be approximated as a square in (3+1)-dimension. There are three gapless modes and one gapped mode.

In (1+1)-dimension, the effective theory in the regime of $ |g_e| > |g_z| $  is a novel realization of a sine-Gordon model coupled with a free scalar field, on which we have made a  renormalization analysis. Described by Kosterlitz-Thouless equations, the phase space  is further divided into three sectors, as shown in Fig.~\ref{fig2},  according to a scaling dimension $t \equiv\frac{1}{\pi}
[\frac{2(g_e^2-g_z^2)}{\xi_+(|g_e|\eta_+ + g_z)}]^{1/2}\kappa(0) -2$  and  a dimensionless parameter $y= \frac{\sqrt{8\alpha}}{\kappa(0)}\lambda$, where  $\kappa$   is a correlation function given in (\ref{correlator}),  $\alpha=\int^\infty_0 \frac{dr}{2\pi}r^3\kappa(r)$.     Both $t$ and $y$ depend on the densities, through $\xi_{+} \equiv \frac{1}{2}(\frac{n_a}{m_a}+ \frac{n_b}{m_b})$ and $\lambda \equiv  \frac{|g_e|n_an_b a^2}{2v},$ respectively. Both the excitation gap in the strong coupling regime and the density-dependent phase transition can be observed in experiments. On the theoretical side, it is interesting to make further studies of the model in the framework of bosonization~\cite{giamarchi,g2}.

\acknowledgments

We thank T. Giamarchi for discussions.
This work was supported by the National Science Foundation of China (Grant No. 11074048) and the Ministry of Science and Technology of China (Grant No. 2009CB929204).

\end{document}